\documentclass[aps,superscriptaddress,prl,twocolumn]{revtex4-1}

\usepackage{amsbsy}
\usepackage{amsfonts}
\usepackage{amssymb}
\usepackage{amsmath}
\usepackage{color}
\usepackage{graphicx}
\usepackage{xspace}
\usepackage[normalem]{ulem}

\usepackage{dsfont}
\newcommand{\idop}{\mathds{1}}

\usepackage{numprint} %Print numbers in scientific writing \numprint{5E3}

\newcommand{\bb}[0]{\begin{eqnarray}}
\newcommand{\ee}[0]{\end{eqnarray}}

\newcommand{\ket}[1]{| #1 \rangle}
\newcommand{\bra}[1]{\langle #1 |}
\newcommand{\moy}[1]{\ensuremath{\langle #1\rangle}\xspace}

\begin{document}
%\showthe\linewidth

\title{%Hybrid opto-mechanical noise in the ultra-strong coupling regime
%Quantum noise in a hybrid opto-mechanical system reaching the ultra-strong coupling regime}
%Quantum mechanical oscillator seen through its ultrastrong coupling to a two level system
%Quantum mechanical oscillator studied through a two level system in the ultrastrong coupling regime
%Probing a quantum-mecanical oscillator with a ultra-strongly coupled two level system
Probing the state of a mechanical oscillator with an ultra-strongly coupled quantum emitter}

\author{Cyril Elouard}\email{celouard@ur.rochester.edu}
\affiliation{Department of Physics and Astronomy and Center for Coherence and Quantum Optics, University of Rochester, Rochester, New York 14627, USA}

%\author{Benjamin Pigeau}
\author{Benjamin Besga}
\email{benjamin.besga@ens-lyon.fr}
\affiliation{Universit\'e de Lyon, CNRS, Laboratoire de Physique de l'\'Ecole Normale Sup\'erieure, UMR5672, 46 All\'ee d'Italie, 69364 Lyon, France.}

\author{Alexia Auff\`eves}\email{alexia.auffeves@neel.cnrs.fr}

\affiliation{CNRS and Universit\'e Grenoble Alpes, Institut N\'eel, F-38042 Grenoble, France}

\begin{abstract}
Performing accurate position measurements of a mechanical resonator by coupling it to some optically driven quantum emitter is an important challenge for quantum sensing and metrology. We fully characterize the quantum noise associated to this measurement process, by deriving master equations for the coupled emitter and the resonator valid in the ultra-strong coupling regime. At short timescales, we show that this noise sets a fundamental limit to the readout sensitivity and that the standard quantum limit can be recovered for realistic experimental conditions. At long timescales, the scattering of the mechanical quadratures leads to the decoupling of the emitter from the driving light, switching off the noise source. This method can be used to describe the interaction of any quantum system strongly coupled to a finite size reservoir.
\end{abstract}

\pacs{Valid PACS appear here}

\maketitle

Coupling optical and mechanical degrees of freedom was first investigated in the pioneering field of cavity opto-mechanics, involving an electromagnetic resonator with a moving end-mirror coupled to a mechanical oscillator \cite{Braginsky1967}. Optical drive of such systems has lead to efficient cooling of the mechanics \cite{Gigan2006,Arcizet2006} down to the ground state \cite{Teufel2011,Chan2011}, opening the path to the manipulation of the quantum states of a macroscopic oscillator \cite{OConnell2010,Palomaki2013}. With the recent developments of nano-mechanics, a new class of systems has emerged where the opto-mechanical coupling is not mediated by a cavity, but by a single quantum emitter \cite{Treutlein2012}. These hybrid systems are now implemented in a wide range of platforms coupling for example single spins \cite{Rugar2004}, Nitrogen-Vacancy (NV) centers in diamonds \cite{Arcizet2011,Teissier2014} or semiconductor quantum dots \cite{Yeo2014,Montinaro2014} to vibrating nanowires, or else in the microwave domain involving superconducting qubits embedded in oscillating membranes \cite{LaHaye2009,Pirkkalainen2013}. Interesting testbeds to investigate the quantum-classical boundary or information thermodynamics \cite{Elouard2014}, these devices are also especially appealing for quantum sensing and metrology \cite{Kurizki2015}. Indeed, tiny variations in the position of the mechanics allow measuring ultra-low forces, as the one created by a single spin \cite{Rugar2004}. Reciprocally, it is possible to extract information on the position of the mechanics from the properties of the light radiated by an embedded quantum emitter \cite{Mercier2015,Munsch2017}. The intrinsic sensitivity of these devices can be enhanced by reducing the size of the mechanical resonator, or alternatively, by increasing the coupling between the mechanics and the quantum emitter: Recently some experiments have reached the ultra-strong coupling regime where the emitter-mechanical coupling is comparable to the mechanical frequency \cite{Yeo2014,Pirkkalainen2013}.

These achievements have opened the way to the experimental study of the ``single-photon regime'' of opto-mechanics ruled by the non-linearized opto-mechanical interaction \cite{Nunnenkamp}, and characterized by new noise sources whose proper modeling is still to come. So far indeed, most theoretical investigations of hybrid opto-mechanical systems have focused on the weak coupling regime \cite{Rabl2010,Wilson-Rae2004,Wang2009,Jaehne2008, QuantumHammer} where the coupling to the quantum emitter is treated using perturbative techniques. A better understanding of state-of-the-art experimental devices now requires to model their evolution in the ultra-strong coupling regime. 

In this letter, we model the dynamics of a mechanical resonator interacting with an optically driven quantum emitter in the ultra-strong coupling regime. We demonstrate that it is possible to extract information on the mechanical position through the light radiated by the quantum emitter. This measurement process induces a back-action noise on the mechanical state, associated to the emitter's population fluctuations. To fully characterize this quantum noise, we go beyond the semi-classical approximation and derive master equations ruling the evolution of the hybrid system. At short timescales, the quantum noise translates into a non-symmetric scattering of the mechanical quadratures, setting a fundamental limit to the sensing precision. We show that the standard quantum limit can be recovered for realistic experimental conditions. At long timescales, this scattering leads to the decoupling of the quantum emitter from the driving light, switching off the noise source. In this picture, the driven quantum emitter behaves both as a measurement channel and as a source of dissipation for the mechanics. Our general method allows describing an unusual situation of quantum optics involving an effective finite size reservoir (the quantum emitter), whose dynamics is sensitive to the evolution of the quantum system (the mechanical resonator). \\

\begin{figure}[t]
\begin{center}
\vspace{0.3cm}
\includegraphics[width=\linewidth]{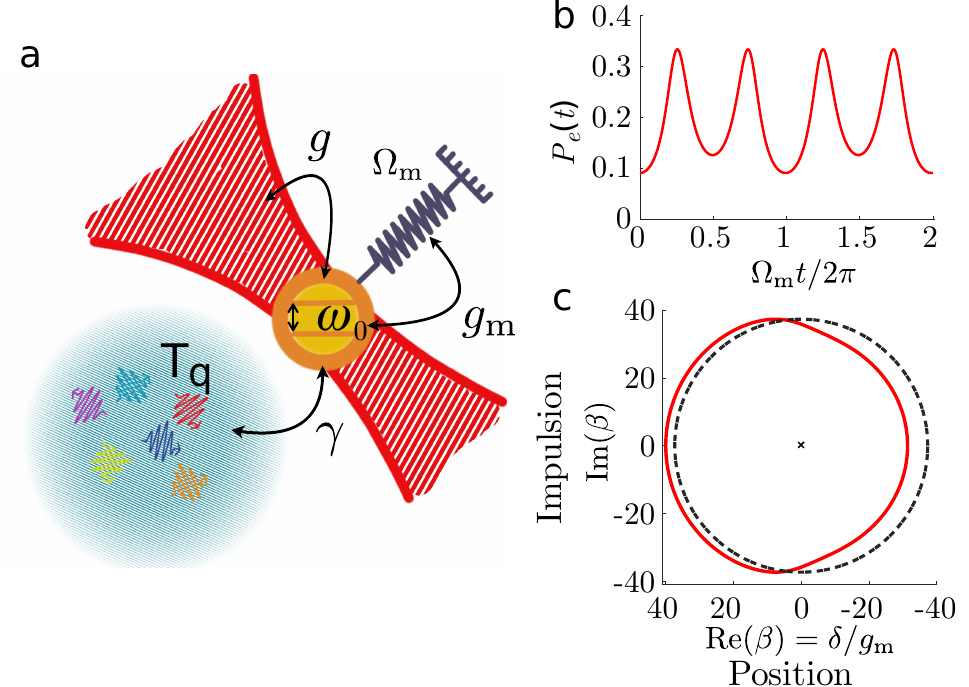}
\end{center}
\caption{\textbf{a}: Hybrid system under study. A two level system of transition frequency $\omega_0$ is optically driven (Rabi frequency $g$), interacts with a heat bath (spontaneous emission rate $\gamma$) and is parametrically coupled to a MO (frequency $\Omega_\text{m}$) with a strength $g_\text{m}$. The coupling is taken adiabatic and ultra-strong $\Omega_\text{m} \leq g_\text{m}\ll \gamma \ll \omega_0$.
\textbf{b,c}: Semi-classical evolution of the hybrid system (cf. equations \eqref{semicl-q} and \eqref{average-m}). \textbf{b}: Modulation of the TLS population $P_e(t) = \bra{e}\rho_\text{q}(t)\ket{e}$. \textbf{c}: Shift of the mean position of the MO in the phase plane (Mechanical complex amplitude $\beta(t) = \text{Tr}\{b\rho^0_\text{m}(t)\}$). Simulation parameters: $\Omega_\text{m}/\gamma = \numprint{5E-3}$, $g_\text{m}/\gamma = 0.1$, $g/\gamma = 1$, $T_\text{q} = 0$. The initial MO state is a coherent state of amplitude $\beta(0) = -20$.}
\label{f:fig1}\end{figure}

\textit{System.---} The hybrid mechanical system under study is depicted in Fig.\ref{f:fig1}a: a two-level system (TLS) of frequency $\omega_0$, of ground and excited states $\ket{g}$ and $\ket{e}$ respectively, is parametrically coupled to a mechanical oscillator (MO) of frequency $\Omega_\text{m}\ll \omega_0$ and driven quasi-resonantly by a classical monochromatic light source of frequency $\omega_L = \omega_0-\delta_0$ where $\delta_0$ stands for the drive-TLS detuning. The Hamiltonian of the driven TLS in the Rotating Wave Approximation is $H_\text{q}(t) = \hbar\omega_0 \Pi_e + \hbar g (e^{i\omega_Lt}\sigma_-+e^{-i\omega_Lt}\sigma_+)$ with $\Pi_e = \ket{e}\bra{e}$ the projector on the TLS excited state, $g$ the classical Rabi frequency, and $\sigma_\pm$ the rising/lowering TLS operators. The MO dynamics is governed by the Hamiltonian $H_\text{m} = \hbar\Omega_\text{m} \left(b^\dagger b + \frac{1}{2}\right)$ and the TLS-MO parametric coupling is $H_\text{c} =\hbar g_\text{m} \Pi_e (b^\dagger + b)$ where $\Omega_\text{m}$ and $b$ are respectively the mechanical frequency and the annihilation operator in the mechanical mode and $g_\text{m}$ is the TLS-mechanical coupling strength. The total Hamiltonian of the hybrid system writes $H(t) = H_\text{q}(t) + H_\text{m} + H_\text{c}$. In what follows, we focus on the ultra-strong coupling regime defined by $g_\text{m} \geq \Omega_\text{m}$. In addition, the TLS is coupled to an electromagnetic heat bath at thermal equilibrium of temperature $T_\text{q}$ and correlation time $\tau_\text{q}$, with  $\gamma$ being the spontaneous emission rate of the bare TLS (reached for $g_\text{m}=0$). We consider here the adiabatic limit $\gamma \gg g_\mathrm{m}$ which is fully compatible with the ultra-strong coupling condition as long as $\Omega_\text{m} \ll \gamma$. \\

\textit{Semi-classical description.---} We assume that at time $t_0$, the hybrid system is in the factorized state characterized by the density matrix $\rho(t_0) = \rho_\text{q}(t_0)\otimes \ket{\phi_\text{m}(t_0)}\bra{\phi_\text{m}(t_0)}$ where $\rho_\text{q}(t)$ is the reduced density matrix of the TLS and $\ket{\phi_\text{m}(t)}$ the (pure) state of the MO.  We first focus on the dynamics of the hybrid system over a few emission and absorption events after time $t_0$.

We choose a coarse graining time $\Delta t_\text{q}$, verifying $\gamma^{-1}\gg \Delta t_\text{q} \gg \tau_\text{q}$ such that the bath degrees of freedom can be traced out \cite{CCT}. As $\Delta t_\text{q} \ll \Omega_\text{m}^{-1},g_\text{m}^{-1}$, the MO remains in the pure state $\ket{\phi_\text{m}(t)}$ and acts as a classical external operator on the TLS, shifting its frequency by $\delta_\text{m}(t) = g_\text{m} \bra{\phi_\text{m}(t)} b+b^\dagger \ket{\phi_\text{m}(t)}$  \cite{Lounis2013,QuantumHammer,Elouard2014} (see Supplemental \cite{SI} Section I). We consider here a mechanical state $\ket{\phi_\text{m}(t)}$  with a well defined position i.e. the quantum variance $V_X = \bra{\phi_\text{m}(t)}X^2\ket{\phi_\text{m}(t)}-\bra{\phi_\text{m}(t)}X\ket{\phi_\text{m}(t)}^2$ of the MO position operator $X = x_0(b+b^\dagger)$, $x_0$ being the zero point motion, fulfills:  
\bb
V_X \ll \dfrac{\gamma}{g_\text{m}}x_0^2.\label{SmallVx}
\ee
This condition ensures that the shift $\delta_m$ of the TLS frequency takes a well-defined value. The master equation for the total hybrid system reads:

\begin{align}  
\dot{\rho}(t) =  -\tfrac{i}{\hbar}[H(t), \rho(t)] +  \left({\mathcal{L}}^{\delta(t)}_\text{q} \otimes \idop_\text{m}\right)[\rho(t)],\label{semicl-tot}
\end{align}

\noindent $\delta(t) = \delta_0+\delta_\text{m}(t)$ is the total detuning between the drive and the TLS, $\idop_\text{m}$ the identity super-operator in the MO Hilbert space. We have introduced the Lindbladian due to the coupling to the heat bath:
\bb
\mathcal{L}^{\delta}_\text{q} = \gamma (n^{\delta}_\text{q}+1)D[\sigma_-] + \gamma n^{\delta}_\text{q} D[\sigma_+], \label{Ldelta}
\ee
where $D[\hat A] \rho = \hat A\rho \hat A^\dagger - \tfrac{1}2 \hat A^\dagger \hat A\rho - \tfrac{1}2 \rho \hat A^\dagger \hat A$ for any operator $\hat A$ and $n^{\delta}_\text{q} = (e^{\hbar(\omega_L+\delta)/k_B T_\text{q}}-1)^{-1}$ is the mean number of thermal photons. Note that we have neglected the modifications of Lindbladian \eqref{Ldelta} due to the drive, which are solely noticeable for very strong driving $\sqrt{g^2+\delta^2}\gtrsim \omega_0$, very high temperature $k_B T_\text{q} \sim \hbar\omega_0$, or structured vacuum \cite{Murch} (see \cite{SI} Section II).  The reduced master equation for the TLS is straightforwardly derived:

\begin{align}  
\dot{\rho}_\text{q} =  -\tfrac{i}{\hbar}[H_\text{q}(t)+ V_\text{q}(t), \rho_\text{q}(t)] +  \mathcal{L}^{\delta(t)}_\text{q}\rho_\text{q}(t)\label{semicl-q},
\end{align}
\noindent where we have introduced the effective Hamiltonian $V_\text{q}(t) = \hbar \delta_\text{m}(t) \Pi_e$.

Eq.\eqref{semicl-tot} is our starting point to investigate the dynamics of the MO over many fluorescence cycles of typical duration $\gamma^{-1}$. The essence of our approach can be grasped, by noticing that the MO-TLS coupling $g_\text{m}$ is much lower than the TLS spectral width $\gamma$: hence the TLS behaves like an effective Markovian reservoir of typical correlation time $\gamma^{-1}$, such that a master equation for the reduced mechanical density matrix $\rho_\text{m}(t)$ can be derived. We thus choose a coarse graining time $\Delta t_\text{m}$ verifying $\gamma^{-1} \ll \Delta t_\text{m} \ll \gamma g_\text{m}^{-2},\Omega_\text{m}^{-1}$. On this timescale, quantum correlations between the TLS and the MO build up and vanish, such that the density matrix of the hybrid system is always factorized $\rho(t) = \rho_\text{m}(t)\otimes \rho_\text{q}(t)$, even in the ultra-strong coupling case. At first order in $g_\text{m}$, the mechanical evolution is governed by the Von Neumann equation:
\begin{align}
 \label{average-m}  
\dot{\rho}^0_\text{m} =  -\tfrac{i}{\hbar}[H_\text{m}+ V_\text{m}(t), \rho^0_\text{m}(t)].
\end{align}
\noindent $V_\text{m}(t) = \hbar g_\text{m} P_e(t) (b+b^\dagger)$ describes the action of the TLS on the MO and $P_e(t) = \mathrm{Tr}[\rho_\mathrm{q}(t) \Pi_e]$ is the mean excitation of the TLS.\\

Eq. \eqref{semicl-q} and \eqref{average-m} are the semi-classical equations of the hybrid opto-mechanical system. The corresponding evolution is pictured in Fig.\ref{f:fig1}b \& c in the case where the MO is initially in a coherent state. On the one hand, the TLS exerts a force on the MO, which translates into a shift of the mechanical rest position. In particular, in the ultra-strong coupling regime a single excitation in the quantum emitter creates a measurable displacement (i.e. larger than $x_0$ the zero point fluctuations of the MO). This hybrid force is a fully analogous to the radiation pressure force in cavity mediated opto-mechanical setups. 

Reciprocally, the MO modulates the frequency of the TLS by $\delta^0_\text{m}(t) = g_\text{m} \text{Tr}[\rho_m^0(t) (b+b^\dagger)]$, varying the value of the total detuning  $\delta(t) = \delta^0_\text{m}(t) + \delta_0$ between the TLS and the driving laser. In the adiabatic regime considered here, this creates a modulation of the TLS population which follows the steady-state solution of optical Bloch equations $P_e^\infty(\delta(t)) = (2+(2\delta(t)/g)^2+(\gamma/g)^2)^{-1}$. Therefore measuring the TLS population, e.g. by recording the intensity of the radiated light can be used to measure the mechanical position. This measurement process is associated to a noise of quantum origin that we shall now characterize.  \\

\begin{figure}%[t]
\begin{center}
\vspace{0.3cm}
\includegraphics[width=\linewidth]{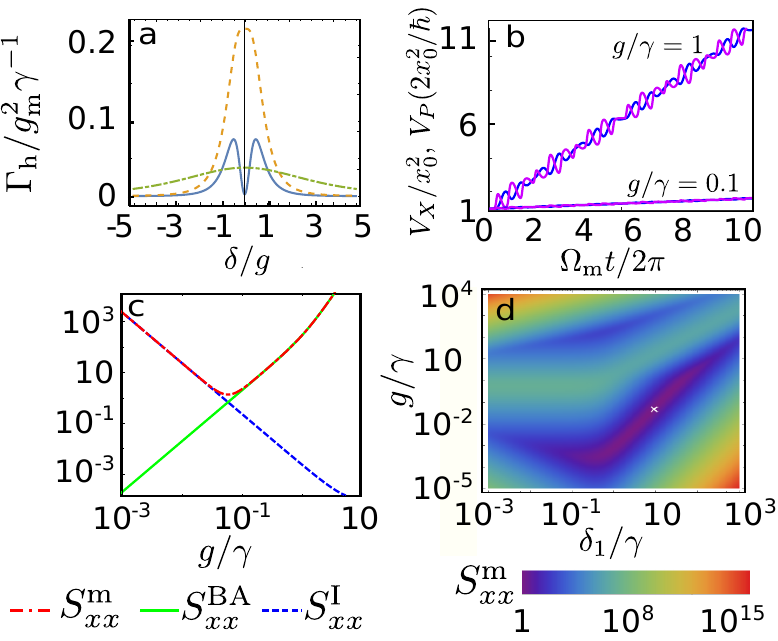}
\end{center} 
\caption{MO position measurement through TLS fluorescence.  \textbf{a}: Hybrid heating rate $\Gamma_\text{h}(t)$ induced by the TLS on the MO  (Eq.\eqref{Gh_adiab}) as a function of the total detuning $\delta(t)$ between the TLS and the driving field for three different Rabi frequencies $g = 10\gamma$ (blue solid), $g=\gamma$ (orange dashed) and $g=0.1\gamma$ (green dash-dotted). \textbf{b}: Evolution of the variances of the MO's position $V_X$ and momentum $V_P$ along $10$ mechanical periods for $g/\gamma = 1$ (top) and $g/\gamma = 0.1$ (bottom). The other parameters for \textbf{a,b} are the same as Fig.\ref{f:fig1}. \textbf{c}: Total measurement noise spectrum $S_{xx}^\text{m}$ and the two contributions: the back-action and imprecision noise spectra $S_{xx}^\text{BA}$ and $S_{xx}^\text{I}$ at the mechanical frequency, for the optimum value of the detuning $\delta_1 \simeq 9.3\gamma$ allowing to reach the global minimum of the total noise, as a function of the Rabi frequency. \textbf{d}: Sensitivity of the MO position measurement as a function of the detuning $\delta_\text{1}$ and the Rabi frequency $g$ in units of $\gamma$. The spectra in \textbf{c,d} are normalized to the minimum value of $S_{xx}^\text{m}$, i.e. $\numprint{4.01E-29}\text{m}^2/\text{Hz}$, reached for $\delta_1 \simeq 8.8\gamma$ and $g\simeq 0.06\gamma$, which is just above the standard quantum limit for the measurement of the position of an oscillator $2x_0^2/\Gamma_\text{m} = \numprint{4E-29}\text{m}^2/\text{Hz}$ \cite{Clerk}. Parameters: $\Omega_\text{m} = 5\text{MHz}$, $g_\text{m}=0.1\text{GHz}$, $\gamma = 1\text{GHz}$, $\Gamma_\text{m} = \Omega_\text{m}/10^6$, $T_\text{q} = 0$ and $x_0 = 10^{-2}\text{pm}$.}
 \label{f:fig2}\end{figure}

\textit{Back-action noise induced by the TLS.---}
We now focus on the evolution of the hybrid system on timescales larger than $\gamma  g_\text{m}^{-2}$, for which the semi-classical description presented above is not valid anymore as the noise induced by the TLS on the MO [i.e. terms of second order in $g_\text{m}$] starts to play a noticeable role. Our strategy is inspired by the projection method of Zwanzig \cite{Zwanzig1961}, which was initially developed to describe the effect of standard reservoirs involving a large number of degrees of freedom. We have adapted this method to the present situation where the spectrally broad TLS plays the role of a ``finite-size reservoir'', sensitive to the evolution of the MO. By coarse-graining the evolution of the hybrid system over a timescale $\Delta t_\text{m} \gg \gamma^{-1}$, we have derived the reduced mechanical equation of motion (See \cite{SI} Section III):

\bb
\dot \rho_\text{m}(t) = -\dfrac{i}{\hbar}[H_\text{m} + V_\text{m}(t),\rho_\text{m}(t)] + {\cal L}_\text{h}[\rho_\text{m}(t)],\label{MmNoBath}
\ee
where
\bb \label{noise}
{\cal L}_\text{h}[\rho_m] = - \dfrac{\Gamma_\text{h}(t)}{2 x_0^2}[X,[X,\rho_\text{m}]].
\ee
\noindent The rate $\Gamma_\text{h}(t)$ is related to the fluctuation spectrum of the TLS population at zero frequency, and is defined as:
\bb
\Gamma_\text{h}(t) = 2 g_\text{m}^2 \text{Re} \int_0^\infty d\tau \moy{\delta \tilde{\Pi}_e(t) \delta \tilde{\Pi}_e(t-\tau)}.
\ee 
We have introduced $\delta \Pi_e(t)= \Pi_e - P_e(t)$ and the tilde stands for the interaction representation with respect to the semi-classical evolution (see Supplemental \cite{SI} Section III.1).
Such a Linbladian has no effect on the average position $X$ or momentum $P = \frac{\hbar}{2x_0} i(b^\dagger - b)$ of the MO, but it increases the momentum variance $V_P (t) = \mathrm{Tr}_\text{m}[\rho_\text{m}(t)P^2]  - \mathrm{Tr}_\text{m}[\rho_\text{m}(t)P ]^2$, leading to a non-symmetric scattering of the mechanical quadratures. Eq.\eqref{noise} describes nothing but the back-action noise of the continuous weak measurement on the mechanical position encoded in the fluorescence light \cite{Jacobs06}. In the adiabatic regime considered here $\Gamma_\text{h}(t)$ verifies (see \cite{SI} Section IV):
\bb
\Gamma_\text{h}(t) = \dfrac{g_\text{m}^2}\gamma \dfrac{2g^2(4\delta(t)^2 + \gamma^2)(g^2 + 2\gamma^2)}{(4\delta(t)^2 + 2g^2 + \gamma^2)^3}. \label{Gh_adiab}
\ee 

Note that $\Gamma_\text{h}>0$, such that the momentum variance can only increase: Therefore no cooling can be realized by driving the TLS. The rate $\Gamma_\text{h}$ characterizing the back-action noise is plotted in Fig. \ref{f:fig2}a. Taking into account the intrinsic coupling of the MO to its own thermal bath with temperature $T_\text{m}$, the full master equation governing the mechanical evolution reads:
\begin{align}
&\dot \rho_\text{m} = -\tfrac{i}{\hbar}\left[H_\text{m}+V_\text{m}(t),\rho_\text{m}\right] + {\cal L}_\text{h}[\rho_m] \nonumber\\
&\quad\quad+ \Gamma_\text{m} (n_m +1) D[ b]\rho_\text{m} + \Gamma_\text{m} n_m D[b^\dagger]\rho_\text{m}.\label{Mm}
\end{align}
We have introduced $\Gamma_\text{m}$ the MO damping rate in the bath, and $n_\text{m} = (e^{\hbar\Omega_\text{m}/k_B T_\text{m}}-1)^{-1}$ the mean number of thermal phonons. 
The weak coupling regime of small $g_\text{m}$ corresponds to $\Gamma_\text{m} \gg \Gamma_\text{h}$: The TLS slightly modifies the effective bath parameters, but the nature of the relaxation remains thermal \cite{Rabl2010} (see \cite{SI} Section V). 
More interestingly, Eq.(\ref{Mm}) allows describing a less explored situation characterized by $\Gamma_\text{h}\sim g_\text{m}^2/\gamma \gg (n_\text{m}+1)\Gamma_\text{m}$ where the TLS-induced mechanical fluctuations largely overcome the Brownian motion.  Such scattering stems from the emission and absorption of photons by the driven dissipative TLS \cite{CCT}. An experimental characterization of this hybrid noise is within reach \cite{Arcizet2011,Yeo2014,Munsch2017}.\\

Eventually, this quantum noise sets a fundamental limit to the sensitivity of measurements based on the optical detection of the MO position \cite{Mercier2017,Mercier2018}, just like radiation pressure in cavity opto-mechanics \cite{Clerk}. To illustrate this point, we focus on the readout of a static MO deflection $\delta_\text{1}$ using the intensity of the TLS fluorescence. The sensitivity of the measurement scheme is quantified by the noise spectrum $S_{xx}^\text{m} $ which is the sum of two contributions: $S_{xx}^\text{m} $ = $S_{xx}^\text{I}$ +$S_{xx}^\text{BA}$  (see \cite{SI} Section VI for analytical expressions). The imprecision noise spectrum $S_{xx}^\text{I}$ is associated to intensity fluctuations of the fluorescence and dominates at low driving (see Fig.\ref{f:fig2}\textbf{c}). At strong driving the quantum noise induced by the TLS $S_{xx}^\text{BA}$ characterized above dominates. The total noise $S_{xx}^\text{m}$  is minimized when both quantum and imprecision noises are equal. $S_{xx}^\text{m}$ is plotted in Fig.\ref{f:fig2}\textbf{d} as a function of the detuning $\delta_1$ and the Rabi frequency $g$. Its minimum value is reached for $(\delta_1/\gamma,g/\gamma) \simeq (8.8,0.06)$ and verifies $S_{xx}^\text{m} = \numprint{4.01E-29}\text{m}^2/\text{Hz}$. This value corresponds to the standard quantum limit for the measurement of the position of an oscillator $2x_0^2/\Gamma_\text{m} = \numprint{4E-29}\text{m}^2/\text{Hz}$ \cite{Clerk}: Thus probing this bound within hybrid opto-mechanical devices is within experimental reach.  \\

\begin{figure}[t]
\begin{center}
\vspace{0.3cm}
\includegraphics[width=\linewidth]{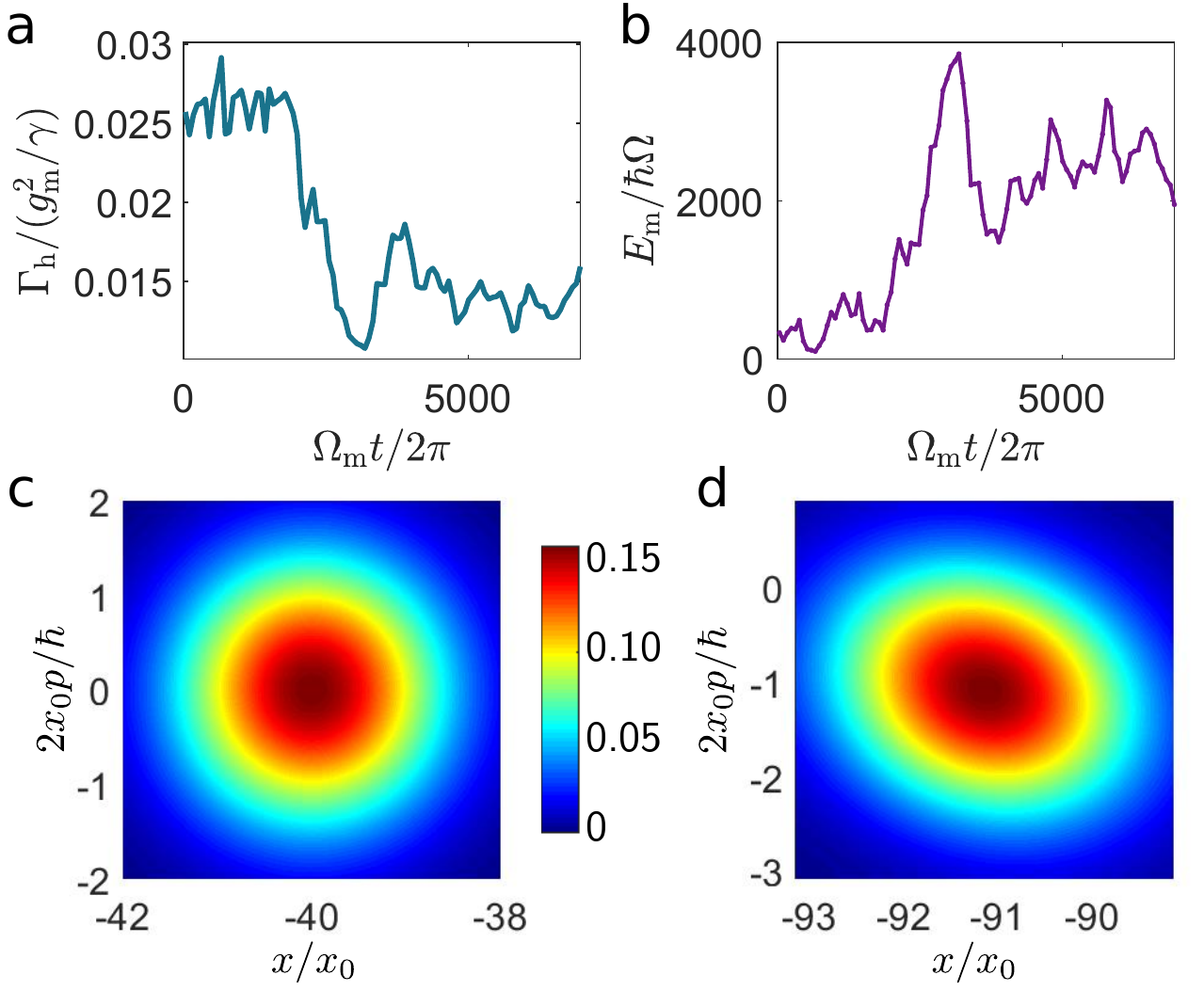}
\end{center}
\caption{\textbf{a,b}: Evolution of the hybrid heating rate $\Gamma_\text{h}(t)$ (\textbf{c}) and of the mechanical energy $E_\text{m}(t) = \hbar\Omega_\text{m}\moy{b^\dagger b}_\text{st}$ (\textbf{d}) along \numprint{3E3} mechanical oscillations. Each point is averaged over $60$ subsequent mechanical oscillations to remove the fast oscillations of the mechanical quantities (see Fig.\ref{f:fig2}). \textbf{c,d}: Wigner function ${\cal W}(x,p)$ of the MO at the initial time when the drive of the TLS is switched on (\textbf{c}) and after \numprint{3E3} mechanical oscillations (\textbf{d}). The simulation parameters are the same as Fig.\ref{f:fig1}.}
\label{f:fig3}\end{figure}

\textit{Quantum trajectory of mechanical oscillator---}  At first sight, it may seem that the momentum variance of the MO and thus the mechanical energy continuously increase under the action the noise, apparently violating the second law of thermodynamics. Actually, over long timescales $t \gg \gamma g_m^{-2}$ the MO position fluctuations randomize the TLS frequency, reducing the coupling to the driving light and eventually switching off the noise source. In this limit the master equation \eqref{Mm} is no longer valid, as Eq.\eqref{SmallVx} breaks down because of the increase of $V_X$. To describe the decoupling, we adopt a quantum trajectory approach taking into account the partial information about the MO position encoded in the radiated light. In this description, at each time $t$ the MO is in a pure state $\ket{{\phi}^\text{st}_\text{m}(t)}$ verifying a stochastic Schr\"odinger equation obtained by unravelling Eq.\eqref{Mm}  \cite{Jacobs06}:

\begin{align}
d\ket{{\phi}^\text{st}_\text{m}(t)}  &=  -\tfrac{idt}{\hbar}\left(H_\text{m}+V_\text{m}(t)\right) \ket{\phi^\text{st}_\text{m}(t)}\nonumber\\
&-\tfrac{dt}{2}\Gamma_\text{h}(t) (X-\moy{X(t)}_\text{st}^2)\ket{\phi^\text{st}_\text{m}(t)}\nonumber\\
& + \sqrt{\Gamma_\text{h}(t)}dW(t)(X-\moy{X(t)}_\text{st})\ket{\phi^\text{st}_\text{m}(t)}\label{Stochm}.
\end{align}
\noindent $dW(t)$ is a real normalized Wiener increment, of zero average and verifying $dW^2(t) = dt$. We have used It\=o's convention for stochastic calculus and denoted $\moy{\cdot}_\text{st}$ the expectation value in state $\ket{{\phi}^\text{st}_\text{m}(t)}$. $\Gamma_\text{h}(t)$ is computed by supposing that the position variance verifies Eq.\eqref{SmallVx} at any time, such that the effective detuning $\delta(t)$ is defined. A trajectory $\ket{{\phi}^\text{st}_\text{m}(t)}$ represents the dynamics of the MO conditioned to a given measurement record $\{X_\text{st}(t)\}_{t_\text{i}\leq t \leq t_\text{f}}$,  of the position measurement performed via the TLS. The measurement outcome at time $t$ is $X_\text{st}(t) = \moy{X(t)} + dW(t)/2\sqrt{\Gamma_\text{h}}dt$ and is stochastic due to the intrinsic randomness of quantum measurement \cite{Jacobs06}. We have solved numerically Eq.\eqref{Stochm} for a single realization of the process. At each time $t$, the scattering rate $\Gamma_\text{h}$ is computed from the instantaneous mechanical state $\ket{\phi^\text{st}_\text{m}(t)}$, generating the trajectory. We find that along each trajectory, the MO follows a random walk in phase space, leading first to the heating of the mechanics (see Fig.~\ref{f:fig3}\textbf{a,b}). Meanwhile, the detuning of the TLS is also scattered, inducing a spectral wandering of the TLS emission line. At longer timescale the mechanical amplitude becomes large compared with $g/ g_m$ and  the MO spends most of the time away from its rest position. The TLS  is brought off resonance with the drive, leading to a vanishing heating rate $\Gamma_\text{h}$ and a saturation of the mechanical energy $E_\text{m}(t)$.

We have plotted the Wigner function ${\cal W}(x,p) = (1/\pi\hbar)\int_{-\infty}^\infty dy \bra{\phi^\text{st}_\text{m}(t)} x+y\rangle\langle x - y\ket{\phi^\text{st}_\text{m}(t)} e^{2ipy/\hbar}$ of the MO at the initial time and after \numprint{3E3} mechanical oscillations (see Fig.~\ref{f:fig3}\textbf{c,d}). One clearly sees the deformation of the shape of the MO state in phase space due to the quadrature-dependent scattering. We also note that the position variance remains of the order of unity along any trajectory (even on large timescales $t\gg \Omega_\text{m}^{-1}$): Indeed, state $\ket{\phi^\text{st}_\text{m}}$ is continuously updated with the information extracted by the continuous position measurement performed by the TLS, which reduces the quantum uncertainty on $X$. Therefore, the effective detuning $\delta(t)$ between the drive and the TLS is defined at any time, validating the trajectory based approach. \\

\textit{Conclusion.---} Our model shows that a hybrid opto-mechanical system in the ultra-strong coupling regime can be described by semi-classical equations at short times, provided that the TLS is strongly dissipative. Beyond the semi-classical regime, the TLS induced mechanical fluctuations either generate an effective thermal bath (small $g_\text{m}$), or the non symmetrical scattering of the MO quadratures (large $g_\text{m}$). This quantum noise is the equivalent of the radiation pressure noise in cavity opto-mechanics, and appears as a fundamental limit of hybrid opto-mechanical detection sensitivity. Noticeable deviations from the semi-classical description can be observed over longer timescales and our study allows to simulate quantum trajectories of the system using a stochastic Schr\"odinger equation, which is a precious tool to describe the TLS and MO fluctuating observables. These quantities are especially relevant to investigate the quantum limit of sensing \cite{Degen2017} in the context of hybrid opto-mechanics, and to perform further thermodynamic studies, e.g. probe fluctuation theorems \cite{CJ}, or design nano-heat engines based on opto-mechanical devices \cite{Kurizki2013,Zhang2014,Dechant2015}. Finally, the method presented here is a quite general one allowing to treat the case of any quantum system in strong interaction with a finite size reservoir \cite{Hu92,Batalhao14,Iles14,Donvil18}.

\begin{acknowledgments}
The authors warmly thank Benjamin Pigeau, Gerard Milburn and Daniel Valente for their advices and support. This work was supported by
the ANR project QDOT (ANR-16-CE09-0010-01). CE aknowledges the support of the US Department of Energy grant No. DE-SC0017890.
\end{acknowledgments}

\pagebreak
\widetext
\begin{center}
\textbf{\large Supplemental Materials: Probing the state of a mechanical oscillator with an ultra-strongly coupled quantum emitter}
\end{center}

%%%%%%%%%% Merge with supplemental materials %%%%%%%%%%
%%%%%%%%%% Prefix a "S" to all equations, figures, tables and reset the counter %%%%%%%%%%
\setcounter{equation}{0}
\setcounter{figure}{0}
\setcounter{table}{0}
\setcounter{page}{1}
\makeatletter
\renewcommand{\theequation}{S\arabic{equation}}
\renewcommand{\thefigure}{S\arabic{figure}}
\renewcommand{\bibnumfmt}[1]{[S#1]}
\renewcommand{\citenumfont}[1]{S#1}
%%%%%%%%%% Prefix a "S" to all equations, figures, tables and reset the counter %%%%%%%%%%

\author{Cyril Elouard}\email{celouard@ur.rochester.edu}
\affiliation{Department of Physics and Astronomy and Center for Coherence and Quantum Optics, University of Rochester, Rochester, New York 14627, USA}

%\author{Benjamin Pigeau}
\author{Benjamin Besga}
\affiliation{CNRS and \'Ecole Normale Sup\'erieure de Lyon, Lyon, France.}

\author{Alexia Auff\`eves}\email{alexia.auffeves@neel.cnrs.fr}

\affiliation{CNRS and Universit\'e Grenoble Alpes, Institut N\'eel, F-38042 Grenoble, France}

\maketitle

\section{Classical behaviour of the MO}

We now consider the total hybrid system (including the MO) coupled to the heat bath. We assume that at time $t$ the hybrid system is in the state $\rho_\text{q}(t) = \rho_\text{q}(t)\otimes\ket{\Phi_\text{m}(t)}\bra{\Phi_\text{m}(t)}$, where $\ket{\Phi_\text{m}(t)}$ is the pure state of the MO. We coarse-grain as in Supplemental I the evolution equation of the total density operator$\rho_\text{tot}^I$ in the interaction picture (which now includes the MO) over the same time $\Delta t_\text{q}$ and trace over the bath degrees of freedom. The interaction picture is now computed with respect to Hamiltonian $H(t) + H_\text{Bq}$ which contains the MO-TLS coupling Hamiltonian $H_\text{c}$ and the MO Hamiltonian $H_\text{m}$. We obtain:

\begin{align}
\Delta\rho_\text{q}^I(t)& = -\dfrac{1}{\hbar^2} \displaystyle\int_t^{t+\Delta t_0}dt' \int_t^{t'}dt''\text{Tr}_\text{Bq}\bigg\{\bigg[ V_\text{Bq}^I(t'), \bigg[ V_\text{Bq}^I(t''),\rho_\text{q}^I(t)\otimes\ket{\phi_\text{m}(t)}\bra{\phi_\text{m}(t)}\otimes \rho_\text{Bq}\bigg]\bigg]\bigg\}.\label{BB:trBq}
\end{align}
where:
\bb
V_\text{Bq}^I(t) = {\cal T} e^{-\tfrac{i}{\hbar}\int_0^t ds (H_\text{q}(s) + H_\text{c})}V_\text{Bq} {\cal T} e^{\tfrac{i}{\hbar}\int_0^t ds (H_\text{q}(s) + H_\text{c})}.
\ee
Here $\cal T$ is the time ordering operator. Because of the presence of $H_\text{c} = (g_\text{m}/x_0) X$ in the interaction picture, the operator $V_\text{Bq}^I(t)$ acts on the MO's Hilbert space. Using the fact that the state $\ket{\phi_\text{m}(t)}$ of the MO present in the integrand of Eq.\eqref{BB:trBq} has a small variance (see condition (1) of main text), we can replace $H_\text{c}$ in $V_\text{Bq}^I(t)$ by its expectation value in that state, namely:
\bb
V_\text{q}(t) = \bra{\phi_\text{m}(t)}H_\text{c}\ket{\phi_\text{m}(t)} = \delta_\text{m}(t)\sigma_z.
\ee 
This allows to apply the end of the derivation of the Lindbladian induced by the heat bath of the TLS taking the value $\delta(t) = \delta_\text{m}(t) + \delta_0$ as the total detuning between the TLS and the drive. Note that the condition (1) of the main text implies that the error maid by taking the average position leads to an error on the detuning of magnitude small with respect to $\gamma$ and which therefore has no influence on the dynamics of the TLS. In particular, it does not affect the population of the TLS given by $P_\text{e}^\infty(\delta(t)) = (2+(2\delta(t)/g)^2 + (\gamma/g)^2)^{-1}$. Assuming the conditions \eqref{AA:flat}-\eqref{AA:approxn} for the Optical Bloch Equations to be valid, we therefore find the master equation (2) of main text.

Finally note that a crucial point for this derivation to hold is that $\delta(t)$ can be considered as constant during the coarse-graining time $\Delta t_\text{q}$ because its typical rates of variation (i.e. the rates of variation of $\ket{\phi_\text{m}(t)}$ are $\Omega,g_\text{m} \ll \gamma \ll \Delta_\text{q}^{-1}$.

\section{Redfield theory for a driven TLS}

In this section we derive the Lindblad equation for the driven TLS weakly coupled to a heat bath at thermal equilibrium, in the absence of coupling to the MO, i.e. $g_\text{m} = 0$. We show that the Standard Bloch Equations \cite{CCT} are valid even at non-zero TLS temperature, and in the saturated regime $g \gg \gamma$. The total Hamiltonian of the TLS writes $H_\text{tot}(t) = H_\text{q}(t) + V_\text{Bq} + H_\text{Bq}$, where $H_\text{Bq}$ is the Hamiltonian of the heat bath, and $V_\text{Bq}$ is the TLS-bath coupling Hamiltonian. We restrict the study to the case $g,\delta_0 \ll \omega_L$, which corresponds to the standard regime of optical Bloch equations. 

The global density matrix in the interaction picture with respect to Hamiltonian $H_\text{Bq}$ and $H_\text{q}(t)$, denoted $\rho_\text{tot}^I(t)$, evolves according to $\dot {\rho}_\text{tot}^I(t) = -\dfrac{i}{\hbar}\left[V_\text{Bq}^I(t),\rho_\text{tot}^I(t)\right]$, where $V_\text{Bq}^I(t)$ is the TLS-environment interaction Hamiltonian in the interaction picture. Following the usual microscopic derivation of Lindblad equation \cite{Breuer,Gardiner2004,CCT}, we average this evolution equation over a time scale $\Delta t_q$, fulfilling $\gamma^{-1} \gg \Delta t_\text{q} \gg \tau_\text{q}$, where $\tau_\text{q}$ is the heat bath correlation time. This leads to: 

\begin{align}
\Delta\rho_\text{q}^I(t)& = -\dfrac{1}{\hbar^2} \displaystyle\int_t^{t+\Delta t_0}dt' \int_t^{t'}dt''\text{Tr}_\text{Bq}\bigg\{\bigg[ V_\text{Bq}^I(t'), \bigg[ V_\text{Bq}^I(t''),\rho_\text{q}^I(t)\otimes \rho_\text{Bq}\bigg]\bigg]\bigg\}.\label{AA:trBq2}
\end{align}
Note that terms of first order in $V_\text{Bq}$ vanish in the trace operation.
We have also performed Born and Markov approximations, i.e. we have replaced the density matrix $\rho_\text{tot}^I(t'')$ in the integrand of \eqref{AA:trBq2} by the factorized density matrix $\rho_\text{q}^I(t)\otimes\rho_\text{Bq}$, where $\rho_\text{q}$ (resp. $\rho_\text{Bq}$) is the reduced density matrix of the TLS (resp. environment). 

At this stage, we are not ensured yet that the right-hand term of equation \eqref{AA:trBq2} is a proper quantum map \cite{Breuer}. We now average over the terms in the integrand oscillating with frequencies larger than $1/\Delta t_0$ (Secular approximation). To do so, we decompose the part of the coupling operator $V_\text{Bq} = R_\text{q}\sigma_x$ acting on the TLS subspace, namely $\sigma_x$, into eigenoperators $$S_\nu(\omega) = \sum_{\epsilon_{s'}- \epsilon_{s} = \omega} \ket{s}\bra{s} \sigma_\nu \ket{s'}\bra{s'},\quad s,s'\in \{+,-\}$$ of the TLS Hamiltonian in the rotating frame $\bar{H}_\text{q} = \delta_0 \sigma_z + \dfrac{g}{2}\sigma_x$ \cite{Breuer,Alicki2013}. $|\pm \rangle = \left( \pm \sqrt{\Omega_R \pm \delta_0} \ket{e} + \sqrt{\Omega_R \mp \delta_0} \ket{g} \right) /\sqrt{2 \Omega_R } $ are the dressed states of the TLS, associated with eigenvalues $\epsilon_{\pm} =\pm \Omega_R/2$, where $\Omega_R = \sqrt{g^2+\delta_0^2}$. 

 In addition, we compute the trace over the environment, forming correlators of the environment operator such that $\moy{R_\text{q}^I(t')R_\text{q}^I(t'-\tau)} = \moy{R_\text{q}(\tau)R_\text{q}(0)}$. These correlators vanish for delay $\tau$ larger than $\tau_\text{q}$, which allows to extend the integral over $\tau$ up to infinity. Defining $\alpha_{\nu}(\omega+\nu \omega_L) = \int_0^\infty d\tau e^{i(\omega+\nu\omega_L) \tau}\moy{R_\text{q}(\tau)R_\text{q}(0)}$, and using $S_\nu(\omega)^\dagger = S_{-\nu}(-\omega)$, we can write:

\begin{align}
\Delta\rho_\text{q}^I(t)=  \sum_{\omega,\omega'}\sum_{\nu,\nu'} \int_t^{t+\Delta t_0}dt' & e^{i(\nu-\nu')\omega_L t'}\alpha_{\nu}(\omega+\nu \omega_L)\nonumber\\
&\times\bigg(S_\nu(\omega)e^{i\omega t'}\rho^I_\text{q}(t)S^\dagger_{\nu'}(\omega')e^{-i\omega' t'} -S^\dagger_{\nu'}(\omega')e^{-i\omega' t'}S_\nu(\omega)e^{i\omega t'}\rho^I_\text{q}(t)\bigg) + h.c.\label{AA:trBq4}
\end{align}

The environment at thermal equilibrium is modeled as a set of independent harmonic oscillators, such as $H_\text{Bq} = \sum_k \hbar\omega_k a_k^\dagger a_k$ and $R_\text{q} = -i\sum_k g_k (a_k^\dagger - a_k^\dagger)$. The corresponding expressions of the bath correlation functions write \cite{Breuer}  $\alpha_+(\omega)= \gamma(\omega)n(\omega)$ and $\alpha_-(-\omega)= \gamma(\omega)(n(\omega)+1)$. We have introduced $\gamma(\omega) = \sum_k g_k \delta_\text{Dirac}(\omega-\omega_k)$ and $n(\omega) = (e^{\hbar\omega/k_b T_\text{q}}-1)^{-1}$, where $\delta_\text{Dirac}$ is the Dirac distribution. From this point, can be derived the Standard Optical Bloch Equations \cite{CCT}, or the Generalized Bloch Equations \cite{Geva1995}, both consistent with $g,\delta_0 \ll \omega_L$. Note that if we want to describe the regime $\Omega_R \gtrsim \omega_L$, we have to choose $\Delta t_0 > \Omega_R$ leading to the Floquet Master Equation \cite{Langemeyer2014,Alicki2013}.

Here, we are interested in the regime of Optical Bloch Equations. We do the following approximations, both motivated by the inequality $\Omega_R \ll \omega_L$: 
\begin{align}
\gamma(\omega_L \pm \Omega_R) &\simeq \gamma(\omega_0) \label{AA:flat}\\
n(\omega_L \pm \Omega_R) &\simeq n(\omega_0) \label{AA:approxn}
\end{align}
for $\nu = \pm$, $\omega = \pm \Omega_R,0$. Approximation \eqref{AA:flat} corresponds to neglecting the variation of the environment spectral density around the frequency $\omega_L$, which is relevant for the electro-magnetic vacuum whose spectral density is proportional to $(\omega_L+\omega)^3$, but may become irrelevant for more complex environment, e.g. when the TLS is within an optical cavity \cite{Murch}. Approximation \eqref{AA:approxn} is valid for $\hbar \omega_L\ll k_B T_\text{q}$ which is the most common case for optically active TLS. This enables to factorize the sums over $\omega$ and $\omega'$ using:
\begin{align}
\sum_\omega \alpha_{\nu}(\omega+\nu\omega_L) S_\nu(\omega)e^{i\omega t'} \simeq  \alpha_\nu(\nu (\omega_L+\delta))\sigma^I_\nu(t')
\end{align} 
As a consequence, when the master equation is written back in Schr\"odinger's picture, all the dependence in time $t'$ of the integrand is contained in the prefactor $e^{i(\nu-\nu')\omega_L t'}$. For $\nu\neq\nu'$, this variation is much faster than the typical evolution time-scale of $\rho_\text{q}^I(t)$, which is given by the inverse of the damping rate of the TLS $\gamma^{-1}(\omega_0) \equiv (v^2\tau_\text{q})^{-1}$. We can therefore safely restrict the sum to the term $\nu = \nu'$ (Secular approximation). This corresponds to a description of the dynamics of $\rho_\text{q}^I(t)$ on a coarse-grained time-scale $\Delta t_0$ much longer than the laser period $\omega_L^{-1}$. Going back to Schr\"odinger picture, we find:
\begin{equation}
\dot \rho_\text{q} = -\dfrac{i}{\hbar}\left[H_\text{q}(t),\rho_\text{q}\right] + \mathcal{L}^{\delta_0}_\text{q}\rho_\text{q}(t)
\end{equation}
with
\begin{equation}
\mathcal{L}^{\delta_0}_\text{q}\rho  = \gamma (n_q+1)D[\sigma_-] + \gamma n_q D[\sigma_+]\label{AA:Lbloch},
\end{equation}
where $\gamma = \gamma(\omega_0)$ and $n_q = n(\omega_0)$. This corresponds to the regime of the so-called (Standard) Optical Bloch Equations \cite{CCT}.

When conditions \eqref{AA:flat}-\eqref{AA:approxn} are no more valid, we have to correct Lindbladian \eqref{AA:Lbloch} to take into account the different terms coming from all the values of $\omega$, which leads to the Generalized Bloch Equations \cite{Geva1995}.

\section{Derivation of the mechanical master equation} 

%\subsection*{1. Zwanzig projection operator method}

In this section we derive the master equation (6) of the main text.  We start from the evolution of the hybrid system (2) of the main text that can be rewritten:
\begin{equation}
\dot{\rho} =  L_t^0[\rho(t)] +  L_t^1[\rho].\label{Eq2}
\end{equation}

We have introduced the semi-classical Liouvillian $L_t^0[\rho] = -\tfrac{i}{\hbar}[H_0(t),\rho(t)] + ({\cal L}_\text{q}^{\delta(t)}\otimes \idop_\text{m})[\rho(t)]$ and $L_t^1[\rho] = -i [\delta  V(t), \rho]$, which creates correlations between the TLS and the MO. We now transform Eq.\eqref{Eq2} to the equivalent of the interaction picture with respect to Liouvillian $L_t^0$. Namely, we introduce the density matrix $\tilde \rho(t)$ verifying $\rho(t) = K(t,0)\tilde\rho(t)$, where $K(t,t_0) = {\cal T}[\exp(\int_{t_0}^t du L_u^0 )]$ is a trace-preserving super-operator and ${\cal T}$ is the time-ordering operator ($K=K_\text{q}\otimes \idop_\text{m}$).
In this picture reflecting the deviation from the semi-classical evolution, the master equation for the hybrid system reads

\bb
 \dot{\tilde{\rho}} = -\tfrac{i}{\hbar}K^{-1}(t,0)[\delta{ V(t)},K(t,0)\tilde{\rho}].
\ee

Defining $\tilde{\rho} = \tilde{\rho}_\text{fact}+\delta\tilde{\rho}$, with $\tilde{\rho}_\text{fact} = \tilde{\rho}_\text{q}\otimes\tilde{\rho}_\text{m}$, we obtain eventually
\bb \label{rho-fact}
\text{Tr}_\text{q} \left\{\dot{\tilde{\rho}}_\text{fact}\right\} = & -\tfrac{i}{\hbar}\text{Tr}_\text{q} \left\{K^{-1}(t,0)[\delta  V(t),K(t,0)\delta \tilde{\rho}]\right\}  \\ \label{delta-rho}
\delta\dot{\tilde{\rho}} = & -\tfrac{i}{\hbar}K^{-1}(t,0)[\delta V(t),K(t,0)\tilde{\rho}_\text{fact}].
\ee

Eq.\eqref{delta-rho} is defined up to first order in $\delta V$. Note that terms in first order in $\delta V$ vanish in \eqref{rho-fact} as they have zero trace over the TLS subspace. Eq. \eqref{delta-rho} is formally integrated and injected into \eqref{rho-fact}, to give an expression close to the familiar precursor for master equation: 

\bb \label{precursor}
\Delta\tilde{\rho}_\text{m} = - \tfrac{1}{\hbar^2}\int_t^{t+\Delta t_\text{m}} dt' \int_t^{t'}dt''\text{Tr}_\text{q}\bigg\{\big[\delta  V(t'), K(t',t'')\big[ \delta  V(t'') , K(t'',0)  \tilde{\rho}_\text{fact}(t'') \big]\big]\bigg\},
\ee
where $\Delta \tilde{\rho}_\text{m} = \text{Tr}_\text{q}\{\tilde{\rho}_\text{fact}(t+\Delta t_\text{m}) - \tilde{\rho}_\text{fact}(t)\}$. To obtain \eqref{precursor}, we have used $K(t',0)K^{-1}(t'',0) = K(t',t'')$. To derive a master equation acting on the MO, we trace over the TLS degrees of freedom. We obtain integrands of the form 

\bb
G_\text{q}(t',t'') \delta \tilde{X}(t')\delta \tilde{X}(t'')\tilde{\rho}_\text{m}(t'') \label{integrand},
\ee

 where we have defined the operators in the interaction picture $\tilde Y(t) = K^{-1}(t,0)Y K(t,0)$ for any operator $Y$ acting on the opto-mechanical system, and the two-times correlation function of the TLS-induced noise:

\bb
 G_\text{q}(t',t'') = \moy{\delta \tilde{\Pi}_e(t')\delta \tilde{\Pi}_e(t'')}.\ee
 
Because of the damping induced by the TLS bath, $G_\text{q}(t',t'')$ vanishes when $\vert t'-t''\vert$ is larger than a typical correlation time $\gamma^{-1}$. As we have chosen $\Delta t_\text{m} \gg \gamma^{-1}$, we can introduce in \eqref{precursor} the variable $\tau = t'-t''$ and extend its interval of integration to $[0;+\infty]$. 

We now consider the second part of the integrand term \eqref{integrand}: $\delta\tilde{X}(t')\delta \tilde{X}(t'')\bar{\rho}_\text{m}(t'')$. This expression can simply be rewritten $(\delta b e^{-i\Omega_\text{m} t'}+ \delta b^\dagger e^{i\Omega_\text{m} t'})(\delta b e^{-i\Omega_\text{m} (t'-\tau)}+ \delta b^\dagger e^{i\Omega_\text{m} (t'-\tau)})\tilde\rho_\text{m}(t'')$, where we have defined $\delta b = b - \text{Tr}\{\rho_m^0 b\}$. The terms proportional to $e^{i \Omega_\text{m} \tau}$ (resp. $e^{-i \Omega_\text{m} \tau}$) can be gathered to perform the integration over $\tau$. 
We define the TLS fluctuation spectrum:

\bb
S_t(\omega) = g_\text{m}^2\int_0^\infty d\tau G_\text{q}(t,t-\tau)e^{i\omega\tau}.
\ee

 Remembering that $\Omega_\text{m} \ll \gamma$ and that $G_\text{q}(t,t-\tau)$ is zero for $\tau \gg \gamma^{-1}$, we can approximate $S_t(\pm\Omega)$ by $S_t(0)$. Taking now into account all the terms from the double commutator in \eqref{precursor}, and coming back to the Schr\"odinger picture, we eventually find:
\begin{align}
&\dot \rho_\text{m} = -\tfrac{i}{\hbar}\left[H_\text{m}+V_\text{m}(t),\rho_\text{m}\right] +  \Gamma_\text{h}(t) D[X]\rho_\text{m},\label{dotrhom}
\end{align}
where we have defined $\Gamma_\text{h}(t) =  2\text{Re}S_{t}(0)$. 

If the MO is also coupled to a thermal bath, the same derivation allows to derive the usual Lindbladian term ${\cal L}_\text{m}[\rho_\text{m}] = \Gamma_\text{m}(n_m +1) D[ b]\rho_\text{m} + \Gamma_\text{m}n_m D[b^\dagger]\rho_\text{m}$ \cite{Carmichael}  that has to be added to Eq.\eqref{dotrhom} 

Finally we get Eq.(9) of main text: 

\begin{align}
&\dot \rho_\text{m} = -\tfrac{i}{\hbar}\left[H_\text{m}+V_\text{m}(t),\rho_\text{m}\right] + {\cal L}_\text{h}[\rho_m] \nonumber\\
&\quad\quad+ \mathcal{L}_\text{m}(t)[\rho_\text{m}]
\end{align}
where $\Gamma$ is the damping rate of the MO, $n_\text{m}$ the number of thermal phonons, and ${\cal L}_\text{h}[\rho_\text{m}] = \Gamma_\text{h}(t) D[X]\rho_\text{m}$ the Lindbladian induced by the TLS on the MO.

\section{Analytical expression of $\Gamma_\text{h}$}\label{s:Spec}

In this section we simplify the expression of $\Gamma_\text{h}$ assuming that the state of the TLS is at any time the steady state of Eq.(3) of main text, namely:

\begin{equation}
\rho_\text{q}^{\infty}= \left(\begin{array}{cc}
P_e^\infty & e^{-i\omega_0 t}(s^{\infty})^* \\ 
e^{i\omega_0 t}s^{\infty} & 1-P_e^\infty
\end{array} \right)
\end{equation}
where
\bb
P_e^{\infty}(\delta(t)) &=& \dfrac{1}{2n_\text{q}+1}\left(n_\text{q} + \dfrac{1/2}{1+2\delta(t)^2/g^2+\gamma_T^2/2g^2}\right)\nonumber\\
s^{\infty}(\delta(t))  &=& -\dfrac{1}{2n_\text{q}+1}\dfrac{\delta(t)/g + i \gamma_T/g}{1+ 2\delta^2(t)/g^2+ \gamma_T^2/2g^2}\label{steadystate}
\ee

We have defined $\gamma_T = \gamma(2n_\text{q}+1)$. The semi-classical evolution of the Hybrid system as described by Eqs.(2)-(3) of the main text implies the following evolution equation for the fluctuation of TLS observables:
\bb
\dfrac{d}{d\tau} \delta\vec{\sigma}(t) = A(t)\delta\vec{\sigma}(t),
\ee
where we have defined the vector $\vec\sigma(t) = (\sigma_z(t),\sigma_+(t),\sigma_-(t))^T$, where $\sigma_r$ for $r\in\{z,+,-\}$ Pauli matrices, and $\delta\vec{\sigma}(t) = \vec\sigma(t) - \moy{\vec\sigma(t)}$.

$$A(t) = \left(\begin{array}{ccc}
-\gamma_T & -ig & ig \\ 
-ig/2 & i\delta(t)-\gamma_T/2 & 0 \\ 
ig/2 & 0 & -i\delta(t)-\gamma_T/2
\end{array}\right).$$

As a consequence, the spectrum
\begin{equation}
S_t(0) = g_\text{m}^2\int_0^\infty d\tau \moy{\delta \tilde{\Pi}_e(t)\delta\tilde{\Pi}_e(t-\tau)}.
\end{equation}
can be computed using the Quantum Regression Theorem \cite{Carmichael}:
\begin{align}
S_t(0) &= -\dfrac{g_\text{m}^2}{4}(1,0,0).A(t)^{-1}\vec G(t),
\end{align}

with

\bb
 \vec G(t) = \left(\begin{array}{c}
1-(\moy{\sigma_z(t)}^\infty)^2\\ 
-\moy{\sigma_+(t)}^\infty(1+\moy{\sigma_z(t)}^\infty)) \\ 
\moy{\sigma_-(t)}^\infty(1-\moy{\sigma_z(t)}^\infty))
\end{array} \right).
\ee

Then, the rate $\Gamma_\text{h}$ is twice the real part of $S_t(0)$. We consider two interesting limits. First, the case of zero TLS temperature $T_\text{q}$:
\begin{equation}
\Gamma_\text{h} = 2\dfrac{g_\text{m}^2 g^2(4\delta(t)^2 + \gamma^2)(g^2 + 2\gamma^2)}{\gamma(4\delta(t)^2 + 2g^2 + \gamma^2)^3}.
\end{equation}

Second, the case of no laser drive ($g=0$) and non-zero TLS temperature:
\begin{equation}
\Gamma_\text{h} = 2\dfrac{g_\text{m}^2}{\gamma}\dfrac{n_\text{q}(1+n_\text{q})}{(2 n_\text{q}+1)^2},
\end{equation}
 Note that $\Gamma_\text{h}$ vanishes if both $T_\text{q}$ and $g$ are zero.

\section{Competition between optical and thermal noise}

\subsection{Diagonal form of total master equation}

 In this section, we introduce another form of the master equation usefull to study the competition between the optical noise induced by the TLS and the thermal noise.
  
The non-unitary part $\mathcal{L}_\text{m}+{\cal L}_\text{h}$ in master equation \eqref{dotrhom} can be written
\begin{align}
\sum_{(i,j)=(1,2)}h_{ij}\left( a_i \rho a_j^\dagger - \dfrac{1}2 a_j^\dagger a_i \rho - \dfrac{1}{2}\rho a_j^\dagger a_i\right).\label{A:Lij}
\end{align}
We have defined the matrix $h=(h_{ij})_{i,j\in\{1,2\}} =  \left(\begin{array}{cc}
 \Gamma_\text{m}(n_\text{m}+1) + \Gamma_\text{h} & \Gamma_\text{h} \\ 
 \Gamma_\text{h} & \Gamma_\text{m}n_\text{m} + \Gamma_\text{h}
 \end{array}\right)$ and the operators $a_1 = b$ and $a_2 = b^\dagger$. This superoperator is reduced to a Lindblad form by diagonalizing the matrix $h$ \cite{Breuer}. We find the eigenvalues $\lambda_\pm$ associated with the eigenvectors $\vec v_\pm$ checking:
 \begin{align}
 \lambda_\pm &= \Gamma_\text{m}\left(n_\text{m}+\dfrac{1}{2}\right)+\Gamma_\text{h}\pm \dfrac{1}2 \sqrt{\Gamma_\text{m}^2 + 4\Gamma_\text{h}^2}\label{C:lpm}
 \\
 \vec v_\pm &= \dfrac{1}{N_\pm}\left(\begin{array}{c}
 \Gamma_\text{m}\pm \sqrt{\Gamma_\text{m}^2 + 4\Gamma_\text{h}^2} \\ 
 2 \Gamma_\text{h}
 \end{array}  \right)\label{C:vpm}
 \end{align}
 with a normalization factor 
 \begin{equation}
 N_\pm = \sqrt{4\Gamma_\text{h}^2 + (\Gamma_\text{m}\pm \sqrt{\Gamma_\text{m}^2 + 4\Gamma_\text{h}^2} )^2}.
 \end{equation}
 The master equation \eqref{dotrhom} can therefore be rewritten:
 
 \begin{align}
&\dot \rho_\text{m} = -\tfrac{i}{\hbar}\left[H_\text{m}+V_\text{m}(t),\rho_\text{m}\right] + \lambda_+D[b_+] + \lambda_-D[b_-],
\end{align}
with

 \begin{equation}
 b_\pm = \vec v_\pm\cdot\left(\begin{array}{c}
 b \\ 
 b^\dagger
 \end{array}\right).\label{C:bpm}
 \end{equation}
 
 We now look at the two limits in which either the thermal or optical noise dominates.

\begin{figure*}
\begin{center}
\includegraphics[width=0.9\linewidth]{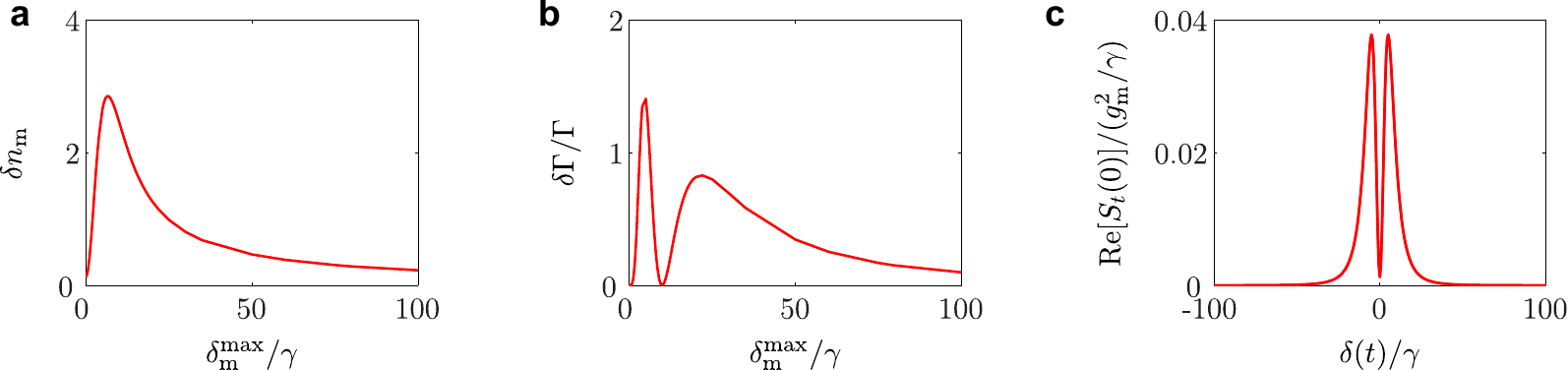}
\end{center}
\caption{\textbf{a,b}: Characteristics of effective thermal noise induced by the TLS  on the MO: Shift of the number of thermal phonons $\delta n_\text{m} = n_\text{m}'-n_\text{m}$ (\textbf{a}) and shift of the damping rate of the MO $\delta \Gamma_\text{m}= \Gamma'-\Gamma$ (\textbf{b}) as a function of the amplitude of the TLS frequency shift  (in $g$ units), i.e. $\delta_\text{m}^\text{max}/g=g_\text{m}\vert \beta\vert_\text{max}/g$. \textbf{c}: real part of the TLS fluctuation spectrum at time $t$ as a function of $\delta(t)$. Parameters: $g/\gamma = 10$, $\Omega/\gamma = 10^{-3}$, $g_\text{m} = 10^{-3}$, $\Gamma_\text{m}/\gamma = 10^{-6}$, $\delta_0 = 0$, $n_m =10^4$, $T_q = 0$.
}
\label{f:Spectra}
\end{figure*}

\subsection*{Weak coupling regime : Effective thermal noise}

In this part we consider the limit $\Gamma_\text{m}\gg \Gamma_\text{h}$. This is in particular valid if $g_\text{m}^2/\gamma \ll \Gamma_\text{m}n_\text{m}$. In that limit, the TLS only slightly changes the non-unitary dynamics induced by the MO environment. We have
 \begin{align}
 &b_+ = b +\dfrac{\Gamma_\text{h}}{\Gamma_\text{m}}b^\dagger\simeq b\\
 &b_- =- \dfrac{\Gamma_\text{h}}{\Gamma_\text{m}a} b+ b^\dagger \simeq e^{-2i\theta} b^\dagger\\
 &\lambda_+ = \Gamma_\text{h} + \Gamma_\text{m}\left(n_\text{m}+1\right) + \dfrac{\Gamma_\text{h}^2}{\Gamma_\text{m}}\\
 &\lambda_- = \Gamma_\text{h} + \Gamma_\text{m}n_\text{m} - \dfrac{\Gamma_\text{h}^2}{\Gamma_\text{m}}.
 \end{align}
The evolution equation for the average phonon number $N = \moy{b^\dagger b}$ writes
\begin{equation}
\dfrac{d}{dt}N = \text{Tr}\{b^\dagger b \dot\rho_\text{m}\} = - \Gamma_\text{m}'(t)(N-n_\text{m}'(t)),
\end{equation}
where we have defined the effective damping rate $\Gamma_\text{m}'(t)$ and the effective MO thermal population $n_\text{m}'(t)$, related to the corresponding parameters in the absence of the TLS:
\begin{align}
\Gamma_\text{m}'(t) &= \Gamma_\text{m}\left(1 + 2\dfrac{\Gamma_\text{h}^2}{\Gamma_\text{m}^2}+o\left(\dfrac{\Gamma_\text{h}^2}{\Gamma_\text{m}^2}\right)\right)\\
n_\text{m}'(t) &= n_\text{m} + \dfrac{\Gamma_\text{h}}{\Gamma_\text{m}} - \dfrac{\Gamma_\text{h}^2}{\Gamma_\text{m}^2}(2 n_\text{m}+1) + o\left(\dfrac{\Gamma_\text{h}^2}{\Gamma_\text{m}^2}\right).\label{nm'}
\end{align}

Under the assumption that the TLS is at any time in the steady state of Eq.(3) of main text corresponding to the instantaneous value of $\delta(t)$, one can find an analytical expression for $\Gamma_\text{h}$ (see following section). The value of $\delta n_\text{m} = n_\text{m}'-n_\text{m}$ and of $\delta \Gamma_\text{m}= \Gamma'-\Gamma$ are plotted in Fig.\ref{f:Spectra} in the case of zero temperature of the TLS and strong driving laser with Rabi frequency $g\gg\gamma$.

\section{Fundamental noise on a position measurement}

We consider a measurement of the position from the intensity of the driven TLS fluorescence in the line of \cite{Munsch2017}. We consider that the MO has some unknown static deflection $x_1$ at time $t=t_1$, resulting in a detuning $\delta_1 = g_\text{m}x_1/x_0$ between the TLS and the drive. The average intensity of the fluorescence reads $E[{\cal I}(t_1)] = \gamma P_e^\infty(\delta_1)$ and allows to infer $x_1$ owing to the dependence of $P_e^\infty$ on the detuning Eq.\eqref{steadystate}. This measurement is subject to two fundamental noise sources \cite{Clerk}: (i) the so-called imprecision noise due to the fluctuation of the light intensity and (ii) the so-called back-action noise induced on the position when the TLS is driven. Both noises can be considered as effective position noises quantified by the spectra $S_{xx}^I(\Omega_\text{m})$ (for the imprecision) and $S_{xx}^{BA}(\Omega_\text{m})$ (for the back-action) taken at the mechanical frequency.\\

The intensity shift $d{\cal I}(t_1)$ associated with a position shift $dx_1$ is $d{\cal I}(t_1) = \dfrac{g_\text{m}}{x_0} {\cal I}'(t_1) dx_1$, where ${\cal I}'(t_1) = \dfrac{d{\cal I}}{d\delta}(t_1) = \gamma \dfrac{dP_e^\infty}{d\delta}(\delta_1)$. Consequently, the imprecision position noise spectrum at a frequency $\omega$ is related to the fluorescence intensity noise spectrum $S_{{\cal II}}(\omega)$ via:
\bb
S_{xx}^I(\omega) = \dfrac{x_0^2}{g_\text{m}^2}\dfrac{S_{{\cal II}}(\omega)}{{\cal I}^{'2}(t_1)}.
\ee
The intensity noise spectrum is defined by
\bb
S_{{\cal II}}(\omega) = 2\text{Re}\int_{0}^\infty  E\left[{\cal I}(t_1+\tau){\cal I}(t_1)\right]e^{i\omega\tau}d\tau,
\ee 
with (see \cite{CCT}[Section $\mathrm{A}_\text{II}$.5], \cite{WisemanBook}[Section 4.3.2]): 
\bb
E[{\cal I}(t_1+\tau){\cal I}(t_1)] = \moy{\sigma_+(t_1)\Pi_e(t_1+\tau)\sigma_-(t_1)}_\infty+ \delta(\tau){\cal I}(t_1),
\ee
where the first term is the (unnormalized) second-order coherence function of the light emitted by the TLS at steady state. The notation $\moy{\cdot}_\infty$ refers in all the section to the average value in the  steady state of the TLS associated with the oscillator's deflection. We finally find for $\omega\neq 0$:
\bb\label{SII}
S_{{\cal II}}(\omega) &=&  2\gamma^2 \text{Re}\left[ \int_{0}^{\infty} \moy{\sigma_+(t_1)\Pi_e(t_1+\tau)\sigma_-(t_1)}_\infty e^{i\omega\tau}d\tau\right] +  {\cal I}(t_1) \nonumber\\
&=& \gamma^2 \text{Re}\left[ \int_{0}^{\infty} \moy{\sigma_+(t_1)\sigma_z(t_1+\tau)\sigma_-(t_1)}_\infty e^{i\omega\tau}d\tau\right] +  {\cal I}(t_1) .
\ee
We then evaluate the first right-hand term in Eq.\eqref{SII}). First, we note that the quantity $\vec Y(\tau) =  \moy{\sigma_+(t_1)\vec\sigma(t_1+\tau)\sigma_-(t_1)}_\infty$, with $\vec\sigma = (\sigma_z,\sigma_+,\sigma_-)^T$, fulfills the vectorial differential equation
\bb
\partial_t \vec Y = A\vec Y + \vec B,
\ee
where $A$ is the matrix defined in Appendix IV (taken for $\delta(t)=\delta_1$) and $\vec B = (-\gamma P_e^\infty(\delta_i),0,0)^T$. This equation can be solved formally leading to:
\bb
\vec Y(\tau) = -A^{-1}\vec B + e^{A\tau}(\vec Y(0)+A^{-1}\vec B).
\ee
Now the first right-hand term in Eq.\eqref{SII} is 
\bb
2\gamma^2 \text{Re}\left[ \int_{0}^{\infty}e^{i\omega\tau} (1,0,0).\vec Y(\tau)d\tau\right]  = 2\gamma^2 \text{Re}\left[ (1,0,0).(i\omega\idop + A)^{-1}.(\vec Y(0)+A^{-1}B)\right].
\ee
Finally, using that

\bb
 \vec Y(0) = \left(\begin{array}{c}
-P_e^\infty(\delta_1)\\ 
0 \\ 
0
\end{array} \right),
\ee
we find that the intensity noise spectrum at the mechanical frequency reads
\bb
S_{xx}^I(\Omega_\text{m}) &=&x_0^2\dfrac{(2\delta_1^2+g^2+\gamma^2/2)^3}{8\delta_1^2  g_\text{m}^2\gamma}\nonumber\\
&&\times\left(\frac{1}{g^2}-\frac{2\gamma^2[3(\gamma^2+4\Omega_\text{m}^2)
-4\delta_1^2]}{\gamma^2(4\delta_1^2+2g^2+\gamma^2)^2+[16(\delta_1^2+g^2)^2+8(g^2-3\delta_1^2)\gamma^2+9\gamma^4]\Omega_\text{m}^2-8[4(\delta_1^2+g^2)-3\gamma^2]\Omega_\text{m}^4+16\Omega_\text{m}^6}\right)\nonumber\\
\ee\\

The back-action noise is related to the noise of the force $F = (\hbar g_\text{m}/x_0) \Pi_e$ exerted on the MO \cite{Clerk}:
\bb
S_{xx}^{BA}(\Omega_\text{m}) &=& \vert\chi(\Omega_\text{m})\vert^2 S_{FF}(\Omega_\text{m})  \nonumber\\
&=& \dfrac{\hbar^2}{x_0^2}\vert\chi(\Omega_\text{m})\vert^2 \Gamma_\text{h},
\ee
where the MO's susceptibility is given by:
\bb
\chi(\omega) = \dfrac{(2x_0^2/\hbar)\Omega_\text{m}}{\Omega_\text{m}^2-\omega^2 - i \Gamma_\text{m}\omega}
\ee

Finally, the total noise added by the measurement reads $S_{xx}^\text{m}(\Omega_\text{m}) = S_{xx}^I(\Omega_\text{m})+S_{xx}^{BA}(\Omega_\text{m})$. The two contributions and $S_{xx}^\text{m}(\Omega_\text{m})$ are plotted in Fig. 2\textbf{c},\textbf{d} of the main text.
Fixing the damping rates of the TLS and the MO to $\gamma = 1\mathrm{GHz}$ and $\Gamma_\text{m} = \Omega_\text{m}/\numprint{1E6} \simeq 5 \textrm{Hz}$ we find that the minimum total noise is reached for $(\delta_1/\gamma,g/\gamma) \simeq (8.9,0.06)$.

\begin{figure*}
\begin{center}
\includegraphics[width=12.5cm]{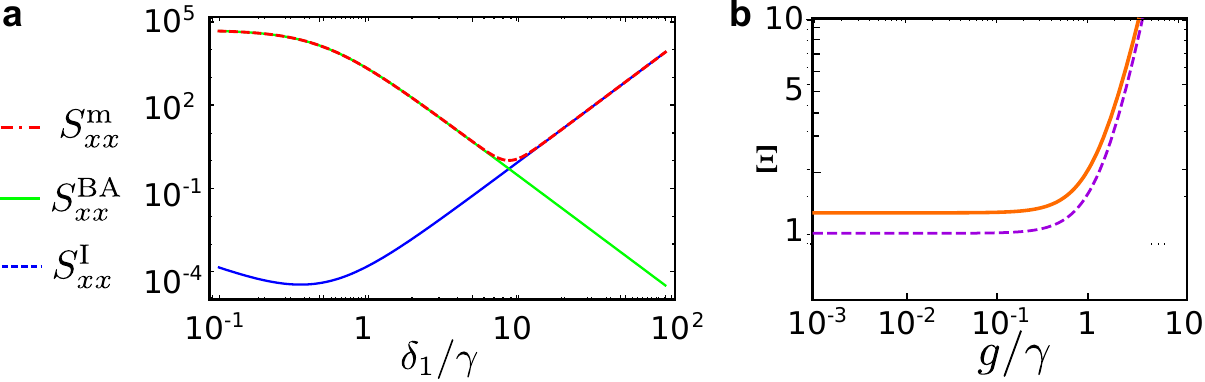}
\end{center}
\caption{\textbf{a}: Total added noise spectrum $S_{xx}^\text{m}$ and the two contributions: the back-action and imprecision noise spectra $S_{xx}^\text{BA}$ and $S_{xx}^\text{I}$ at the mechanical frequency, for the optimum value of the Rabi frequency, i.e. $g\simeq = 0.06\gamma$, as a function of the detuning $\delta_1$ corresponding to the static deflection of the MO. \textbf{b}: Parameter $\Xi$ quantifying the distance to Heisenberg as a function of the drive Rabi frequency, for the optimum value of the detuning $\delta_1 \simeq 8.8\gamma$ (purple dashed) and for $\delta_1 = \gamma$ (orange solid). The equality is approximately reached for the optimum value of detuning. Parameters:  Parameters: $\Omega_\text{m} = 5\text{MHz}$, $g_\text{m}=0.1\text{GHz}$, $\gamma = 1\text{GHz}$, $\Gamma_\text{m} = \Omega_\text{m}/10^6$ and $x_0 = 10^{-2}\text{pm}$, $\gamma = 1$ GHz, $x_0 = 10^{-2}$pm, $\Gamma_\text{m} = \Omega_\text{m}/10^6$, $T_\text{q} = 0$.
}
\label{f:Xi}
\end{figure*}

The two quantities $S_{xx}^I(\Omega_\text{m})$ and $S_FF(\Omega_\text{m})$ are expected to fulfill an Heisenberg inequality of the form \cite{Clerk}:
\bb
S_{xx}^I(\Omega_\text{m})S_{FF}(\Omega_\text{m}) \geq \dfrac{\hbar^2}{4},
\ee
This can be tested by studying the ratio:
\bb
\Xi = \dfrac{S_{xx}^I(\Omega_\text{m}) S_{FF}(\Omega_\text{m})}{\hbar^2/4} \geq 1.
\ee

We find that indeed $\Xi \geq 1$ and that the bound is approximately saturated $\Xi \simeq 1$  in the regime $\delta_i \gg \gamma \gg g$.

\end{document}